\documentclass[a4paper,fleqn,longmktitle]{cas-sc}
\usepackage[authoryear,longnamesfirst]{natbib}
\usepackage{algorithm}
\usepackage{algorithmic}

\usepackage{graphicx}
\usepackage{natbib}
\graphicspath{ {figs/} }
\def\tsc#1{\csdef{#1}{\textsc{\lowercase{#1}}\xspace}}
\tsc{WGM}
\tsc{QE}
\tsc{EP}
\tsc{PMS}
\tsc{BEC}
\tsc{DE}

\begin{document}
\let\WriteBookmarks\relax
\def\floatpagepagefraction{1}
\def\textpagefraction{.001}
\shorttitle{EER: Enterprise Expert Ranking using Employee Reputation}
\shortauthors{Mahmood et,al.}

\title [mode = title]{EER: Enterprise Expert Ranking using Employee Reputation}                      

\tnotetext[1]{This document is the results of the research
   project funded by the National Science Foundation.}

\tnotetext[2]{The second title footnote which is a longer text matter
   to fill through the whole text width and overflow into
   another line in the footnotes area of the first page.}

\author[1,2]{Saba Mahmood}

\ead{saba.mahmood@gmail.com}

\credit{Conceptualization of this study, Methodology, Software}

\address[1]{Department of Computer Science \& Software Engineering, International Islamic University Islamabad, Pakistan}
\address[2]{Department of Computer Science, Bahria University Islamabad, Pakistan}

\author[1]{Anwar Ghani}[style=chinese, orcid=0000-0001-7474-0405]
\cormark[2]
\ead{anwar.ghani@iiu.edu.pk}

\author[3]{Ali Daud}[%
   role=,
   suffix=,
   ]
\cormark[1]
\ead{ali_msdb@hotmail.com}

\credit{Data curation, Writing - Original draft preparation}

\address[3]{Department of Information Systems and Technology, College of Computer Science and Engineering, University of Jeddah, Jeddah, Saudi Arabia }

\author%
[1]
{Syed Muhammad Saqlain}
\ead{syed.saqlain@iiu.edu.pk }


\cortext[cor1]{Corresponding author}
\cortext[cor2]{Principal corresponding author}
\fntext[fn1]{This is the first author footnote. but is common to third
  author as well.}
\fntext[fn2]{Another author footnote, this is a very long footnote and
  it should be a really long footnote. But this footnote is not yet
  sufficiently long enough to make two lines of footnote text.}

\nonumnote{This note has no numbers. In this work we demonstrate $a_b$
  the formation Y\_1 of a new type of polariton on the interface
  between a cuprous oxide slab and a polystyrene micro-sphere placed
  on the slab.
  }

\begin{abstract}
The emergence of online enterprises spread across continents, have given rise to the need of expert identification in this domain. Scenarios that includes the intention of the employer to find tacit expertise and knowledge of an employee that is not documented or self-disclosed has been addressed in this article. The existing reputation based approaches towards expertise ranking in enterprises  utilize pagerank ,normal distribution and hidden markov model for expertise ranking. These models suffer issue of negative referral, collusion ,reputation inflation and dynamism. The authors have however proposed a Bayesian approach utilizing beta probability distribution based reputation model for employee ranking in enterprises. The experimental results reveal improved performance compared to previous techniques in terms of Precision and Mean Average Error (MAE) with almost 7\% improvement in precision on average for the three data sets. The proposed technique is able to differentiate categories of interactions in dynamic context. The results reveal that the technique is independent of the rating pattern and density of data.
\end{abstract}




\maketitle

\section{Introduction}

Web information is available in the form of websites, micro-blogs, email, and other social networking sites.Expert ranking has emerged as a new area of research since information present on the web can be authored by anyone especially in the case of micro-blogs, thus a requirement was felt to rank the expertise level of contributors\cite{expert20a}.Expert is someone with high level of knowledge related to a certain subject ~\cite{6}.Expert ranking is studied in different scenarios like finding an expert in micro-blogs~\cite{1}.Newer areas of expert findings includes author ranking~\cite{2,3,3a} and employee/contractor ranking in a large organization or online job portals~\cite{4}~\cite{expert20b},~\cite{expert20c}Recently finding influencers in bibliometric~\cite{expert20e} networks also come under expert finding techniques in academia. More recently authors~\cite{5} have proposed evaluation of web content credibility through expert ranking.The rise in large online enterprises, expert identification in large organizations and enterprises has emerged. 
Two major approaches towards expert ranking are graph-based and document-based~\cite{7,17}. The graph-based techniques explore the social connections of the author, while the document-based techniques explore the documents produced by the author as evidence of expertise. Researchers have also utilized a hybrid approach to expert ranking~\cite{19}.

 Previous techniques that are graph-based utilizes Pagerank like algorithms for ranking, unfortunately, none of them have addressed the issue of collusion and negative referrals. The document-based techniques relate an author with the topic, techniques like LDA and topic modeling is utilized. Document-based techniques are limited by scenarios where the quality of documents are not available. The problem under discussion in this research is the identification and ranking of experts in large organizations. The objective of expert ranking in enterprise organizations is to answer questions like ``Who is an expert on subject X?'' or ``Does X have knowledge of subject Y?''.Although previous researchers have claimed that document-based techniques are suitable for organization since documents are present in an organized manner, but authors are considering the situation when there is an intention to find the tacit expertise~\cite{expert20d}~\cite{expert20f} of an employee not present in documented form or when management is interested to find expertise level of an employee for an additional task not related to his/ her documented area of expertise. In such scenarios document-based techniques are insufficient, the link/graph-based techniques are also incapable to address this issue, thus in order to address this research gap authors have proposed a reputation based scheme. The reputation is calculated through feedback from other users. The technique is compared to baseline techniques i.e. page rank based expert rank~\cite{7}, worker rank~\cite{8} and expertise assessment in online labor markets~\cite{4} treated as baselines. The experimental results of  precision and mean average error show better performance of the proposed EER technique in comparison to the baselines in identification of the experts.

The paper consists of sections of Research Objectives, Related Work that highlights the literature review, The EER Technique, Problem Formulation, Evaluations, and Conclusion.

\section{Research Objectives}
The expert rank techniques in the context of enterprise and large organizations is rarely researched . However evolution of online enterprises spread across continents has raised importance of these techniques specifically when the skill sets are not documented(tacit).The previous techniques generically had limitations in terms of negative referral. In case of domain under 
 discussion reputation based approaches are inadequate in truly representing the opinions and suffered 
inability of adaptation to  dynamism. The reputation calculation structures are inappropriate to incorporate all kinds of interactions resulting in reputation inflation .Thereby authors highlighted and aimed towards following research objectives:
\begin{itemize}

   \item To design expert rank algorithms for a large organization/enterprises to find the tacit talent of employees.
   \item A reputation-based expert rank technique based on the type of interactions.
   \item A technique that can solve negative referrals and collusion problems of previous graph-based techniques.
   \item To propose a solution that can use the time-based reputation with the ability to judge dynamic behaviors.
\end{itemize}

\section{Related Work}
This section of the paper highlights the literature of expert ranking in different domains with focus on enterprises and large organizations. The authors have further categorized literature according to different techniques utilized in expert identification.

\subsection{Expert Ranking domain category: Enterprise and micro blogs}

Knowledge in organizations and large enterprises is documented and quality of information is high due to disciplined and organized policy of documentation compared to online knowledge communities and blogs \cite{7}.However, authors have claimed expert finding a problem in organizations as discussed by \cite{9} and \cite{10}. Sources of information including self-disclosed information, document and social network based information can provide evidence for expert ranking in enterprises.
Self-disclosed information is hard to attain and update that may require more time, while document based and social network based indicators  are significant that can be automated to find an expert.

Most organizations observe that expert finding is used to find people outside the organizations only~\cite{11},primarily due the assumption that employees within the organization are well known for their expertise and skills. However with the emergence of large enterprises, that are geographically distributed with employees from different knowledge backgrounds, education, skills, merged employees from other organizations expert finding, within the organization has also emerged as an important dimension.In certain scenarios the knowledge of employees is not documented and is thus tacit.

\begin{figure}
\centering
  \includegraphics[width=1.0\linewidth]{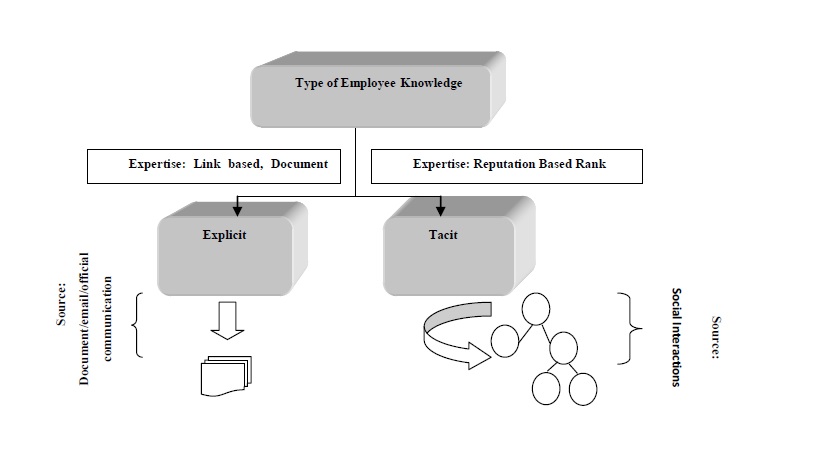}
  \caption{Employee Knowledge Categorization}
  \label{fignew}
\end{figure}

Research in the domain of expert finding falls into two categories, i.e. graph-based and document-based, few researchers have explored the problem domain under discussion however they utilized one of the two approaches or a hybrid one. In case of employee ranking in organizations and enterprises, the techniques depends upon the availability of the type of knowledge regarding employee expertise.If the knowledge regarding employees is tacit and no documentary proof exists, social interactions are the source of information , while link based, document based techniques have been utilized  when explicit knowledge in the form of documentary proof/email communication/official communication exists as shown in the figure~\ref{fignew}.Social interactions have been explored using link and graph based techniques however due to their limitations, reputation based techniques using different mathematical structures have been proposed by the researchers.A recent research workerRank i.e. ranking of workers utilizes reputation information and document-based techniques. Authors\cite{4} of another reputation based technique utilizing the HMM model has also addressed  problem of expert identification in organizations.

\begin{figure}
\centering
  \includegraphics[width=1.0\linewidth]{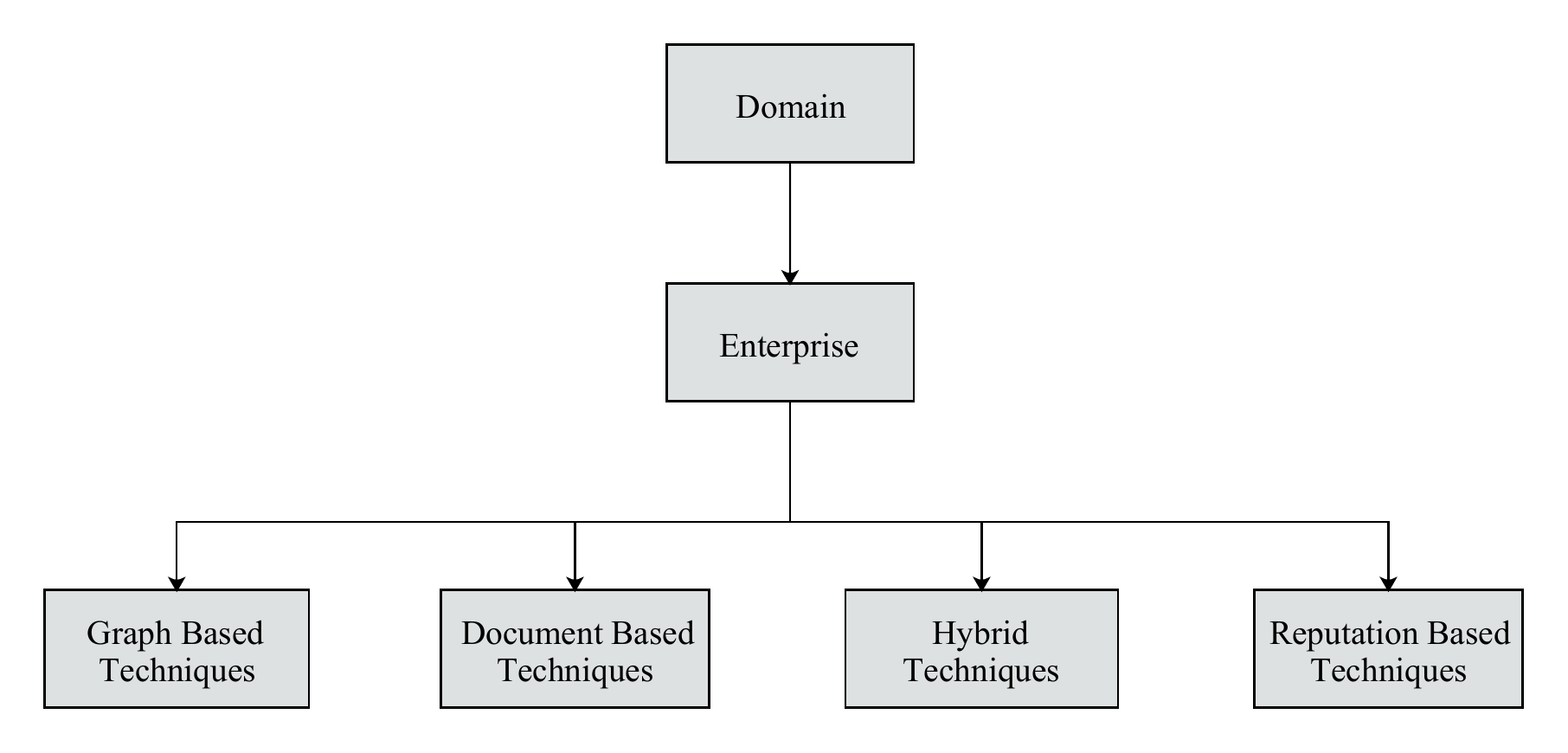}
  \caption{Classification of techniques for expert finding in enterprises}
  \label{fig1a}
\end{figure}

Given below is the detailed literature survey of expert ranking techniques based on social network (graph-based), document-based techniques, hybrid and reputation based techniques, also shown in the figure \ref{fig1a}.

\subsection{Social network/Graph-based Expert Identification Systems}
There has been an increase in the research related to identification of experts.COGNOS~\cite{12}is a technique based upon wisdom of crowds. It utilizes twitter list, to find an expert. The most commonly occurring name in a subject returned by the crawler is considered to be the subject expert. The COGNOS was tested with plenty of experiments that however highlighted a shortcoming against the malicious and fake users.Such users can create fake lists so as to undermine true experts.
The Twitter Who to Follow service~\cite{13} provides query capability to find an expert. It utilizes self disclosed information of users in terms of bio and other details along with its social connections.In a more recent work~\cite{14}the authors carried out a survey that revealed that an advisor can be regarded as reliable and trustworthy through analysis of his review history.
Page rank ~\cite{15} that basically ranks web pages with high rank if it is pointed by popular pages. An advancement of Page Rank algorithm for twitter is the Twitter Rank~\cite{16}. The algorithm finds experts from the number of followers of the tweets. In order to find topic experts they utilized LDA technique to relate influential tweets with a topic.
The authors of this work~\cite{18} are of the view that experts returned by twitter rank are generally well known thus they proposed metrics of number of tweets, followed tweets and  replied tweets. The data returned from these metrics are then clustered by using Gaussian Mix Algorithm. The evaluation was carried out against the survey results.

Expert Rank~\cite{7}is inspired by the PageRank algorithm,finds an expert by analysing the documents produced by the user along with user's influence in the social network.However this technique and others that utilize graph based techniques are unable to differentiate types of interactions. They have targeted online knowledge communities. The algorithm analyzes the documents produced by the candidate expert as well as his value/rank in the community, together they produce better results.
Theses techniques take number of following or number of interactions are however,unable to address number of unfollowings or negative interactions.

Expert Finding through social referrals~\cite{19} finds an experts through referrals being made to neighbors chosen on the basis of profile match and associated cost. If the target is not found among the immediate neighbors, further search is carried out. Upon completion of the search when a desired expert is found, the initiator pays everyone in the referral chain. Divya et al\cite{a1}\cite{a2} utilized email communications of an organization to find experts using link structures. Similarly expertise rank\cite{a6} from question answer links of online courses discussion forum is also based upon graph techniques.

\subsection{Document-based Techniques}
These techniques are utilized to find topic based experts from the documents produced by the candidate experts.These techniques usually employ text mining  to relate topic with an expert. Most popularly the LDA~\cite{17} technique is utilized to relate a topic with the expert. This technique is utilized by twitter rank , that applies the LDA to the tweets and posts produced by the users. This technique is effective when large amount of organized documentation is available also LDA puts all document data under one topic. Research under probabilistic topic modelling\cite{a10} is also document based and utilizes different mechanism for text analysis, that are then related to topic based experts.\cite{a11}  utilizes generalized LDA technique for topic modelling from the social network annotations.

\subsection{Hybrid Systems}
There are certain systems that utilize both document and social network information to find experts. For example Campbell\cite{10} utilized the popular HITS algorithm on the social network being built from the email communication.They also analyzed the content of the email.However the technique was only appropriate for small sized data sets.Hybrid approach \cite{a3}\cite{a4} utilizing subject relationship,and user influence gathered from link analysis are utilized to find experts.Another work \cite{19} utilized  the message thread in the online java forums along with analysis of the content of messages. J Wang et al \cite{a5} used convoluted neural networks to predict user with best answer thus indicating it as subject expert.

\subsection{Reputation-based techniques}

In literature techniques that utilize reputation based information for experts includes work by Faisal et al~\cite{21}that addresses the issue of expert finding in micro-blogs, the calculation structure adopted by the technique is a simple summation. Worker Rank~\cite{8} ranks employees of an organization by utilizing weighted average method to calculate reputation. These weights are generated from normal distribution based calculation structure.More recently a Hidden Markov Model (HMM)~\cite{4}based reputation technique has been proposed for expertise assessment in online labor markets.This work addresses the issue of inflated reputation score and reputation staticity in a dynamic environment. They are of view that the reputation rank must be according to recent interactions. Such issue has been addressed by authors~\cite{22} in time bound topic modeling for expert ranking.Answer reputation\cite{a8}\cite{a9} was measured utilizing number of accepted answers up voted answers. The reputation model is simpler and ignores user consistency and tags\cite{a7}.
Table\ref{t1} summarizes various expert identification techniques along with their methodologies and the parameters required by them to execute the technique.

\begin{table*}
\centering
\caption{Expert Identification Systems Comparison}
\label{t1}
\begin{tabular}{@{}l l l@{}}
\toprule
\textbf{Framework }						& \textbf{Parameters} 					& \textbf{Techniques}\\
\midrule 
COGNOS												& Twitter List 									&	Mining Twitter List\\
Profile History								&	Review History								& Mining Review History\\
Twitter Rank 									&	Followers											&	PageRank, LDA\\
Topical Authorities						&Follow,Reply										& Gaussian Mixture Model\\
Expert Rank										&	Document Analysis, Social Rank&	PageRank\\
Social Referral								&	Social Connections						& Profile Matching\\
Fu et al											&Email communication						& Graph +document approach\\
Hecking et al. & Question/Answers & Graph\\
Divya et al & Email links & Link Structures\\
Worker Rank										&	Workers job/skills						& Document + Reputation (NDR)information\\
Expert in online labor market	&	Skills 												&	Hidden Markov Reputation model\\
\bottomrule
\end{tabular}
\end{table*}

\section{Problem Formulation}

This section defines the basic terminologies used in EER followed by a formal definition of the problem of Expert Employee Ranking.

\textbf{Interactions Categorization}:  Interactions are categorized as alpha or beta, whereby alpha represent all kinds of positive relations, feedback, ratings, for example in case of social media they can be number of followings. While beta represents all kind of negative interactions, relations, feedback, ratings. For example number of unfollowings represent beta in a scenario of social networks for instance.

\textbf{Expert Employee:} An employee is considered as an expert if he possesses highest level of skills pertaining to his subject area while Tacit  Expertise refers to expert with highest level of skills for which there is no documentary proof.

\textbf{Definition (Enterprise Expert ranking using Employee Reputation):} Let $E=\{e_1, e_2, e_3, \cdots, e_n\}$ be the set of expert users and $Ev$ represents the expected value. Let $I$ be the set of all interactions in which user $U$ has participated, where $I=\{i_1, i_2, i_3,\cdots, i_m\}$ and categorization results in $i=\{\alpha/\beta\}$ and $U=\{u_1, u_2, u_3,\cdots,u_n\}$ such that $U_i$, has  some documented and undocumented skills. Authors assume that a user$U_i$ is member of $E$  if he has participated in interaction $I_i$ , if and only  if the  $Rep((Ev), U_i,I)$ is maximum. Where $Rep()$ utilizes beta probability density function to compute the expected value for set $I$ and $U_i$.
 
\section{Expert Employee Reputation (EER)}

The proposed methodology intends to address the following problems
\begin{itemize}
\item Identification of an expert in an organization for which no document exists, where the intention of an employer is, to find tacit expertise of an employee.
\item Finding an employee suitable for an extra task not directly related to the documented qualification and expertise.
\item Existing graph-based techniques for expertise ranking used in the given domain are vulnerable to the issue of negative referrals and collusion.
\end{itemize}

\begin{figure}
\centering
  \includegraphics[width=0.9\linewidth]{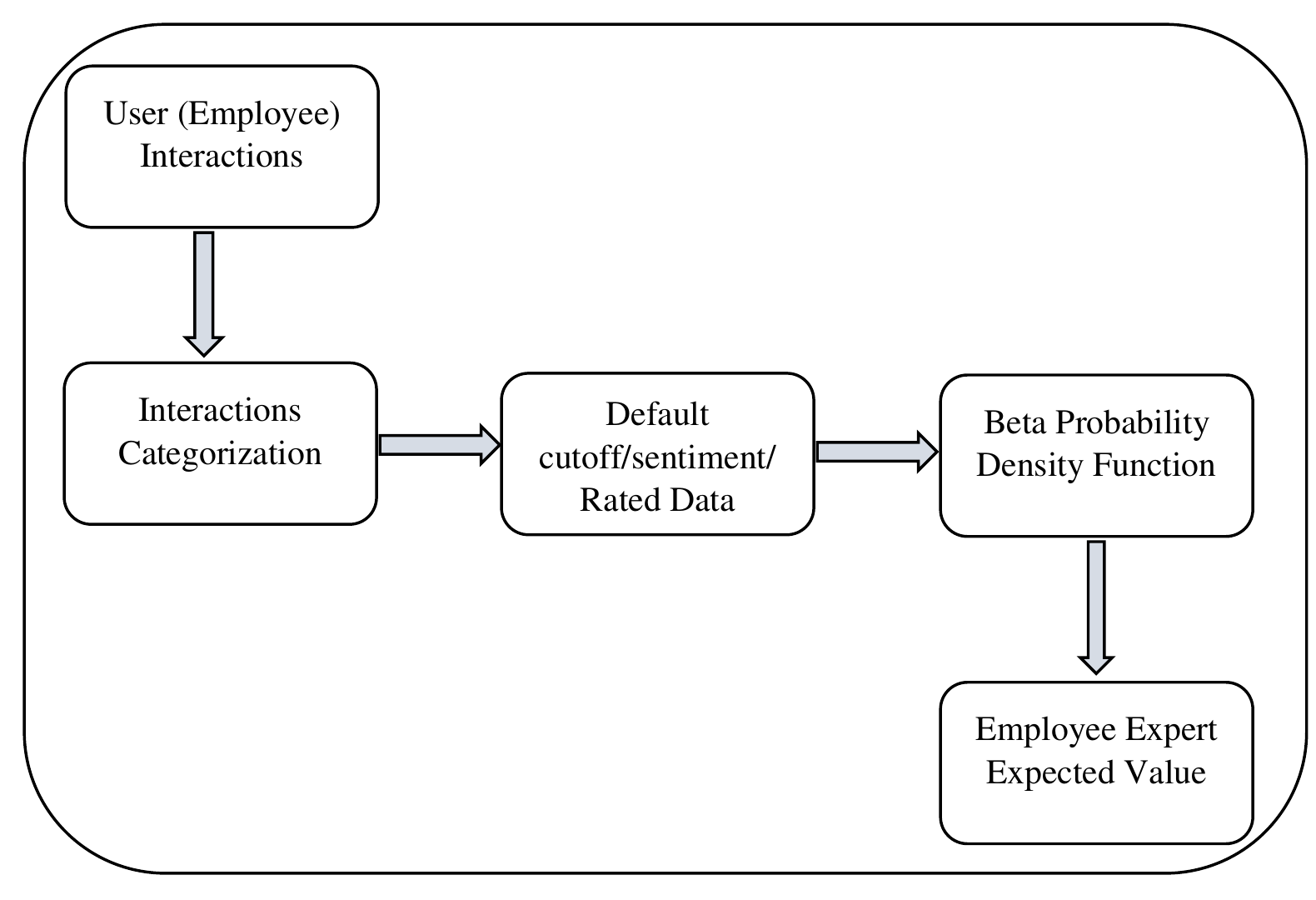}
  \caption{EER Architecture}
  \label{fig1}
\end{figure}

To address these problems authors found that document-based and graph-based techniques will be insufficient so they proposed a reputation-based approach that ranks and identifies an expert through reputation feedback and interactions.
The reputation information can be gathered from direct opinion or through organizational micro-blogs, where an employee might be involved in the discussion are under consideration.Given in Figure~\ref{fig1} is the Architecture of  EER. The employee interactions are categorized according to a criteria, this information is then  given as an input to Beta Probability Density function that generates the expected value of the employee treated as his reputation rank.

Reputation is defined as ``Overall quality as seen and judged by users'' according to Merriam-Webster's~\cite{23} online dictionary. 
The reputation is calculated from the opinion about previous behavior of the entities derived from the history of interactions. 
The past behavior of the entities and opinions of others is utilized to find the reputation. The opinion is based on the history of interactions. However, there are scenarios when information is not present in the form of opinions. In such cases, interactions might be in the form of text messages, comments or posts. The following section presents details of  the two major modules of EER, i.e. interaction categorization and the beta probability density function.

\subsection{Interactions Categorization}

Interactions are the communication, ratings, followings, query answer reply.Interaction categorization is proposed since previous techniques were unable to identify  them  for example in case of twitter rank or others that are based on page rank are not able to take into account the number of unfollowings or number of dislikes.Thus in EER if interactions are present in the form of text, sentiment analysis~\cite{24}\cite{a12} is one such technique that can be utilized to categorize the interactions as either positive or negative. If the interactions are measured in terms of continuous values, a threshold needs to be decided for the binary classification. Thus the values above that threshold are considered as positive and those below are considered as negative.The decision regarding this threshold or cutoff in most of the cases requires ground truth values\cite{25}.Mean Probability and $0.5$ is however satisfactory threshold for interaction categorization.
Sometimes such interactions are also stated explicitly as either positive or negative thus they can be then utilized further directly.

\subsection{Beta Probability Distribution}

Beta probability distribution~\cite{26} finds the posterior probability of the binary events. Beta probability density function is parameterized by alpha and beta representing the two events.In EER alpha and beta are represented by the positive and negative interactions between the employees of the enterprise respectively. Given below in equation\ref{e1a} is the expected value.

\begin{equation}
\label{e1a}
	E(v)= \frac{\alpha}{\alpha +\beta} 
\end{equation}

Beta probability can be used to represent the subjective degree of belief. Let's assume positive interaction between employees and negative interactions in the enterprise as the two events, where positive interactions are represented by alpha and beta represents negative interactions.Assuming $'A'$ represents number of activities of a context, where  $'x'$ represents a particular context. Thus $'A'$ are activities of context $'x'$.
If $'M'$ represents the total number of nodes in the network, the expert node in the context $'x'$ can be computed by utilizing equation~\ref{e1a}. 

 Supposing $z$ and $z_1$ are the number of outcomes of alpha and beta respectively, that implies that after every $z$ outcome the $z_1$ outcome can be expected. In the EER, assume $p$ are the observed number of outcomes for $z$ and $n$ are the observed number of outcomes for $z_1$, then following equations can be derived. 
\begin{equation}
\label{e1}
	\alpha = z + 1
\end{equation}

\begin{equation}
\label{e2}
	\beta = z_1 + 1
\end{equation}

\begin{equation}
\label{e3}
E(v)= \frac{z + 1}{z + z_1 + 2} 
\end{equation}

Substituting the number of outcomes for $z$ and $z_1$, we get

\begin{equation}
\label{e4}
	E(v)= \frac{p + 1}{p + n + 2} 
\end{equation}

The Expert Reputation of a node is the sum of expected value $E(v)$ of all its interactions, given by

\begin{equation}
\label{e5}
	 E=\sum_{i=1}^{M-1} E(v) 
\end{equation}
where $'E'$  called as expected value represents the reputation  of a node in a particular context.

The proposed technique is different in terms that it considers the type of interaction of the users. For every user in the network, the positive and negative interactions are recorded. These are then utilized to find the expected posterior behavior of the nodes by utilizing the beta probability expected value as discussed above.  Based upon these values all the nodes are ranked. The node with maximum reputation value is regarded as the expert node. Given below is the Algorithm \ref{algo-1}, the details are already discussed in the previous section. In Line 3, 4 the interactions are categorized into positive or negative domains. Line 5 uses the beta probability expected value to generate the expert rank of the node.

\begin{algorithm}
\caption{Reputation based Expert Rank}
\label{algo-1}
\begin{algorithmic}[1]
\STATE{Load Dataset}
\LOOP
	\STATE{Compute no of positive interactions p for worker $w$}
	\STATE{Compute no of negative interactions n for worker $w$}
	\STATE{Compute expected value for worker $w$ as}
	\STATE{$Ev \leftarrow p+1/p+n+2$} 
	\STATE{Let M represent total no. of interactions}  
	\LOOP
		\STATE{$E\leftarrow \sum_{i=1}^{M-1} E(v)$}
	\ENDLOOP
	\STATE{Let T represent the total number of nodes} 
	\STATE{Compare the Reputation of worker w with $T-w$ nodes }
	\STATE{$max\leftarrow w$}
\ENDLOOP
\STATE{Compute max as the Expert} 

\end{algorithmic}
\end{algorithm}

Lines 7-10 compute the expert rank of a node for all its interactions. The results are then added. Lines 11-13 compare the expert rank of a node to the rest of nodes in the network. The node with the maximum expert rank value is then regarded as the expert node.
Given in algorithm\ref{algo-2} is the part of the algorithm with the time factor.

\begin{algorithm}
\caption{The Time factor Algorithm}
\label{algo-2}
\begin{algorithmic}[1]
\STATE{Find time t node i interacting node $n-i$}
\IF{$t\leftarrow 0$} 
	\STATE {Latest interaction only node i,$n-i$}
	\ELSE 
		\STATE{All interactions node i,$n-i$}
\ENDIF
\end{algorithmic}
\end{algorithm}

Algorithm~\ref{algo-2} states that if the value of variable $t =0$ then only the latest interaction is counted. Otherwise,a history of interaction can be counted. The algorithm can be fine-tuned to include a particular length of history. Like previous $(1, 2,  ..10)$ or any specific number of interactions in history thereby excluding rest. This feature of the algorithm addresses the dynamism of the expert ranking with the time that was solved using HMM's reputation in one of the baselines.

\section{Evaluation}
Two performance indicators within~\cite{27} this context are reported.The average absolute error \cite{28} between real and predicted.The metric compares the reputation values calculated by the proposed technique against the real values in order to find  ability of the technique in predicting the rankings close to real rankings.Other metric is Precision that is used to find the probability by which the proposed technique truly ranks the expertise.


\subsection{Baselines}
Authors carried out comparison to expert rank~\cite{7} that is a graph based technique utilizing PageRank like algorithm and Worker Rank \cite{8}, a technique utilizing normal distribution based reputation model for ranking employees in organizations according to their expertise.

Furthermore, another recent technique using Hidden Markov Model (HMM) \cite{4} as a reputation calculation structure is also used for evaluation. The HMM-based technique proposed to solve the issue of reputation inflation, and change in reputation value with time  that is not being addressed in other reputation models in general. The proposed beta probability-based reputation model has incorporated the time factor to address this shortcoming. Thus, comparison to this technique is made under the time decay metric.Since the scenario under discussion lacks documentary proof, therefore the comparison against document based techniques stands void.


\subsection{Datasets}

The effectiveness of the reputation-based expert rank algorithm is found by performing experiments on three different datasets~\cite{29}. Table~\ref{t2} highlights these datasets.
 
 \begin{table}
\centering
\caption{Datasets}
\label{t2}
\begin{tabular}{@{}c c@{}}
\toprule
\textbf{Datasets} &  \textbf{Nodes}\\
\midrule 
Dataset DS1  & 46 \\ 
Dataset DS2  & 77\\
Dataset DS3  & 77\\

\bottomrule
\end{tabular}
\end{table}

Dataset DS1:This data set represents the interactions in term of expertise level of the users. The weights are based on a scale from 0 to 5. 0: I Do Not Know This Person; 1: Strongly Disagree; 2: Disagree; 3: Neutral; 4: Agree; and 5: Strongly Agree.

Dataset DS2: This data set represents the ratings given according to the degree of advice received from the users of the network. The scale of the weights is  0: I Do Not Know This Person/I Have Never Met this Person; 1: Very Infrequently; 2: Infrequently; 3: Somewhat Infrequently; 4: Somewhat Frequently; 5: Frequently; and 6: Very Frequently.

Dataset DS3:This third data set is about the employees knowledge regarding the skills and expertise  of each other.
 The weight scale is 0: I Do Not Know This Person/I Have Never Met this Person; 1: Strongly Disagree; 2: Disagree; 3: Somewhat Disagree; 4: Somewhat Agree; 5: Agree; and 6: Strongly Agree.

\subsection{Ranking Match Test}
This test is carried out to find out the capability of the proposed technique(EER) in truly representing the ratings given by the individual nodes.The results are compared against the graph based baseline that utilizes page rank algorithm~\cite{15} referred as BL1 and a reputation based technique for ranking workers in an enterprise~\cite{8} referred as BL2 in text.The reputation structure adopted by this technique is normal distribution(NDR)\cite{30}. Scenario under discussion in this paper is based on the hypothesis of absence or lack of documentary proofs, thus comparison to the third category of document based technique is not carried out.

\begin{figure}
\centering
  \includegraphics[width=0.95\linewidth]{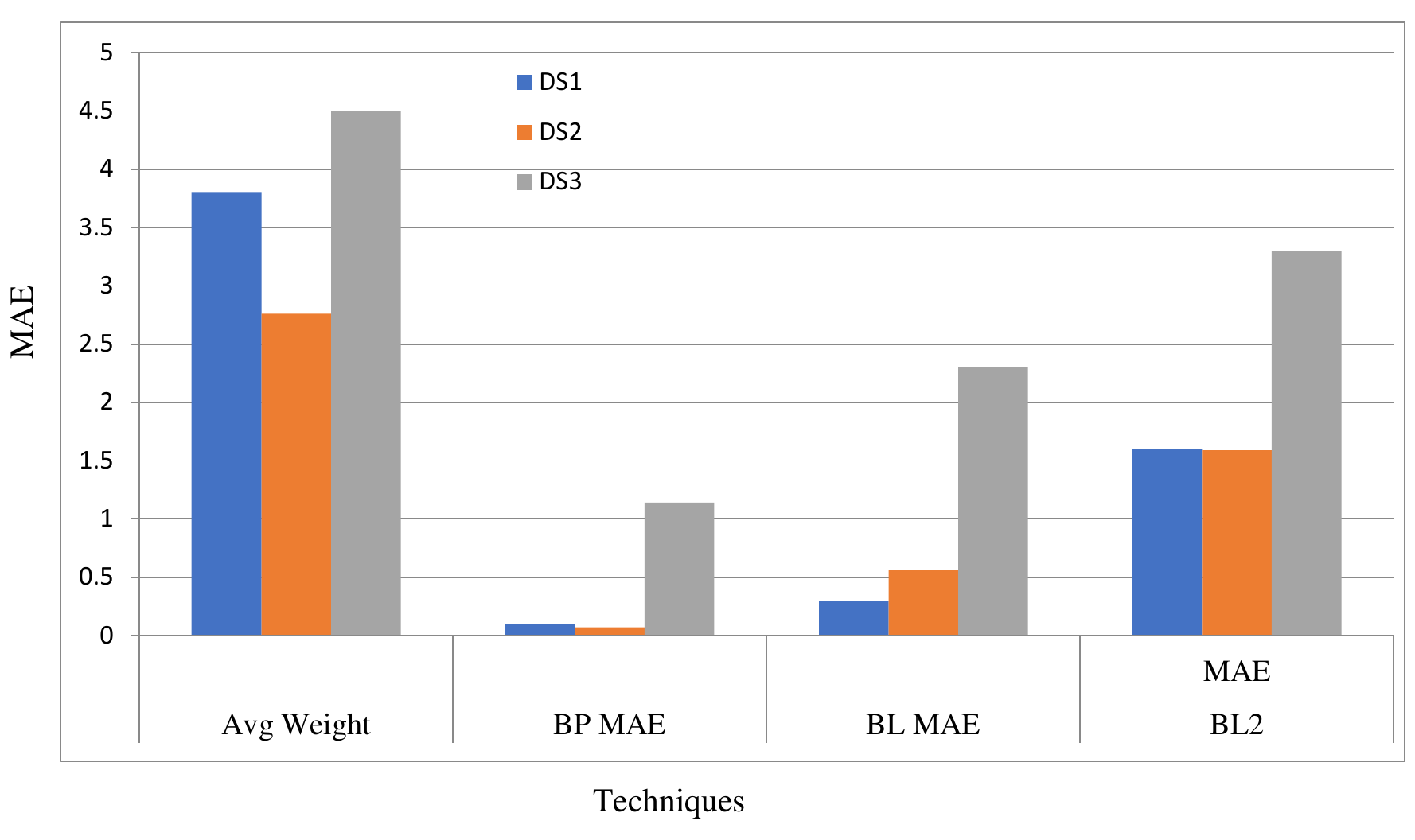}
  \caption{Comparison of EER and Baselines w.r.t to MAE}
  \label{fig2}
\end{figure}

The test was carried out on Datasets DS1, DS2 and DS3. Comparison  of reputation values produced by the proposed technique against the real values is carried out by using the  performance metric  MAE(Mean Average Error). This evaluates the capability of EER in predicting rankings close to real ones.The result is shown in Table~\ref{t3}and Fig~\ref{fig2}.The Mean average Error value for EER technique is 0.1 as compared to 0.3 for BL1 and 1.6 for BL2 in case of DS1. Same trend is found for other two data sets.

 \begin{table}
\centering
\caption{MAE of EER, Baseline Algorithms}
\label{t3}
\begin{tabular}{@{}l c c c c@{}}
\toprule
Data Sets&	Avg Weight&	EER MAE&	BL1 MAE&	BL2MAE\\
\midrule 
DS1    &3.8	& 0.1	& 0.3	&1.6\\
DS2	&2.76&	0.07	&0.56&	1.59\\
DS3	&4.5	&1.14&	2.3	&3.3\\
\bottomrule
\end{tabular}
\end{table}

 \begin{table}
\centering
\caption{P$@$10 for the three datasets}
\label{t4}
\begin{tabular}{@{}l c c c @{}}
\toprule
Datasets&	BL2(Worker Rank)&	BL1&	EER\\
\midrule
DS1	&0.0	&0.0	&0.06\\
DS2	&0.1	&0.15	&0.2\\
DS3	&0.0	&0.1	&0.2\\
\bottomrule
\end{tabular}
\end{table}

The results demonstrate the accuracy of the EER technique in representation of the expert nodes.

\begin{figure}
\centering
  \includegraphics[width=0.95\linewidth]{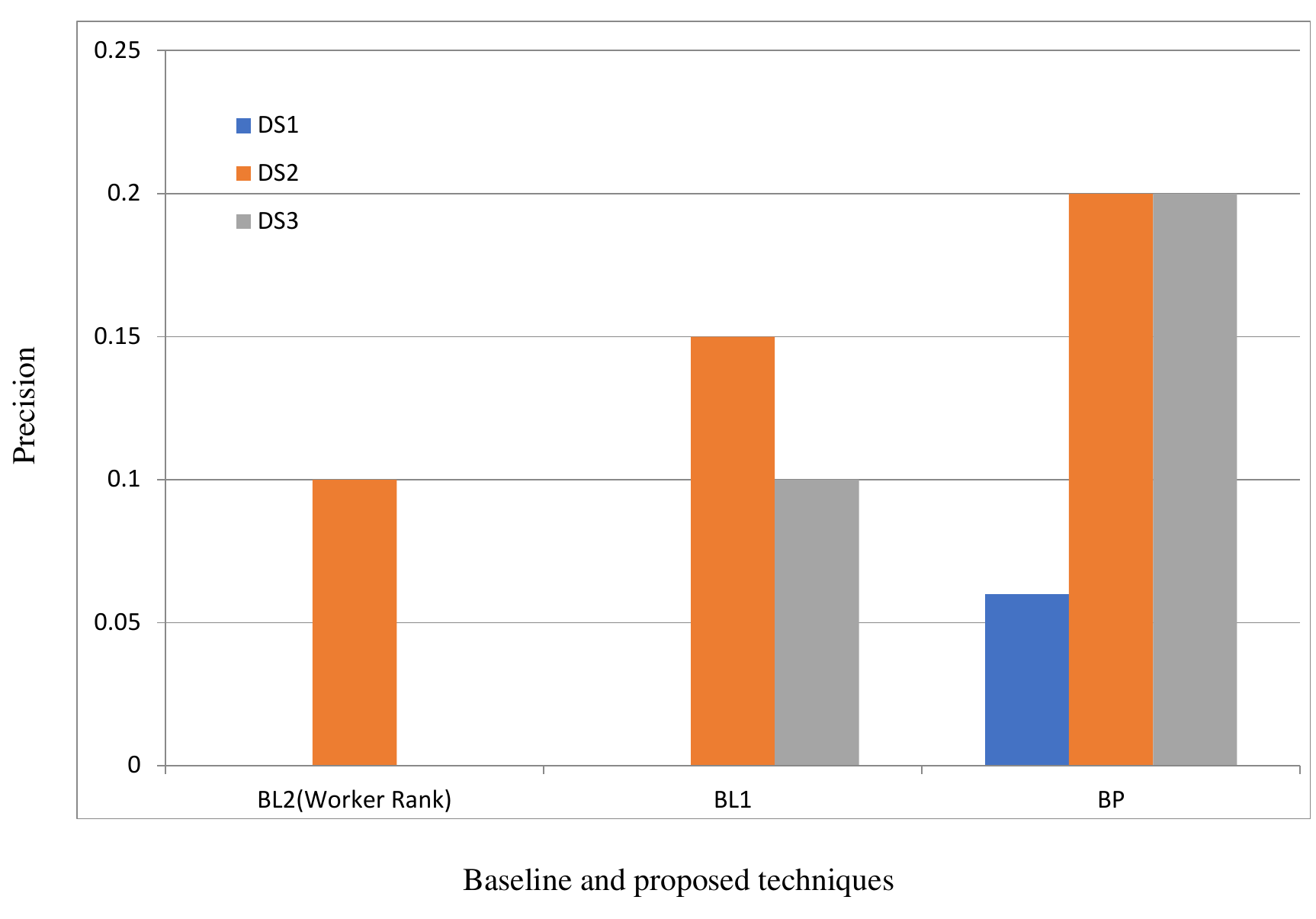}
  \caption{Comparison of techniques w.r.t Precision}
  \label{fig3}
\end{figure}

\textbf{\textit{Analysis}}:
The MAE result from the three datasets reveal that,EER yields lesser value when compared with the baselines as shown in table \ref{t3} and figure \ref{fig2}. Similarly, the precision results in table \ref{t4} and figure\ref{fig3} represent enhanced performance of EER.The precision results of BL2 are nearly zero ,analysis revealed the  reason that the weights in BL2 are normally distributed. Comparatively performance of BL1 is better than BL2; but it also yields zero result for the data set DS1.Further analysis of this outcome revealed that in dataset DS1 every node has equal number of connections.The BL1 page rank based technique is limited by such scenarios, even the weighted page ranks are also unable to support results.By carefully observing the results of the proposed technique against all three datasets it is found that precision for DS2 and DS3 are better compared to DS1, the reason being the density of data. The DS1 is denser compared to the other two datasets. Thus,the proposed technique produces better results for sparse datasets. It is also observed that for DS2 precision of BL2 (worker Rank) is quite promising revealing the fact that the ratings in that dataset are normally distributed as compared to DS1 and DS3. Overall on average for the three datasets the Precision of EER is almost 7\% improved compared to baselines.

This all discussion leads authors to summarize that the proposed technique EER is independent of the pattern of ratings and the density of the datasets whereas other baselines seem to change its behavior with these factors.

\subsection{Interaction Categorization Test}

\begin{figure}
\centering
  \includegraphics[width=0.95\linewidth]{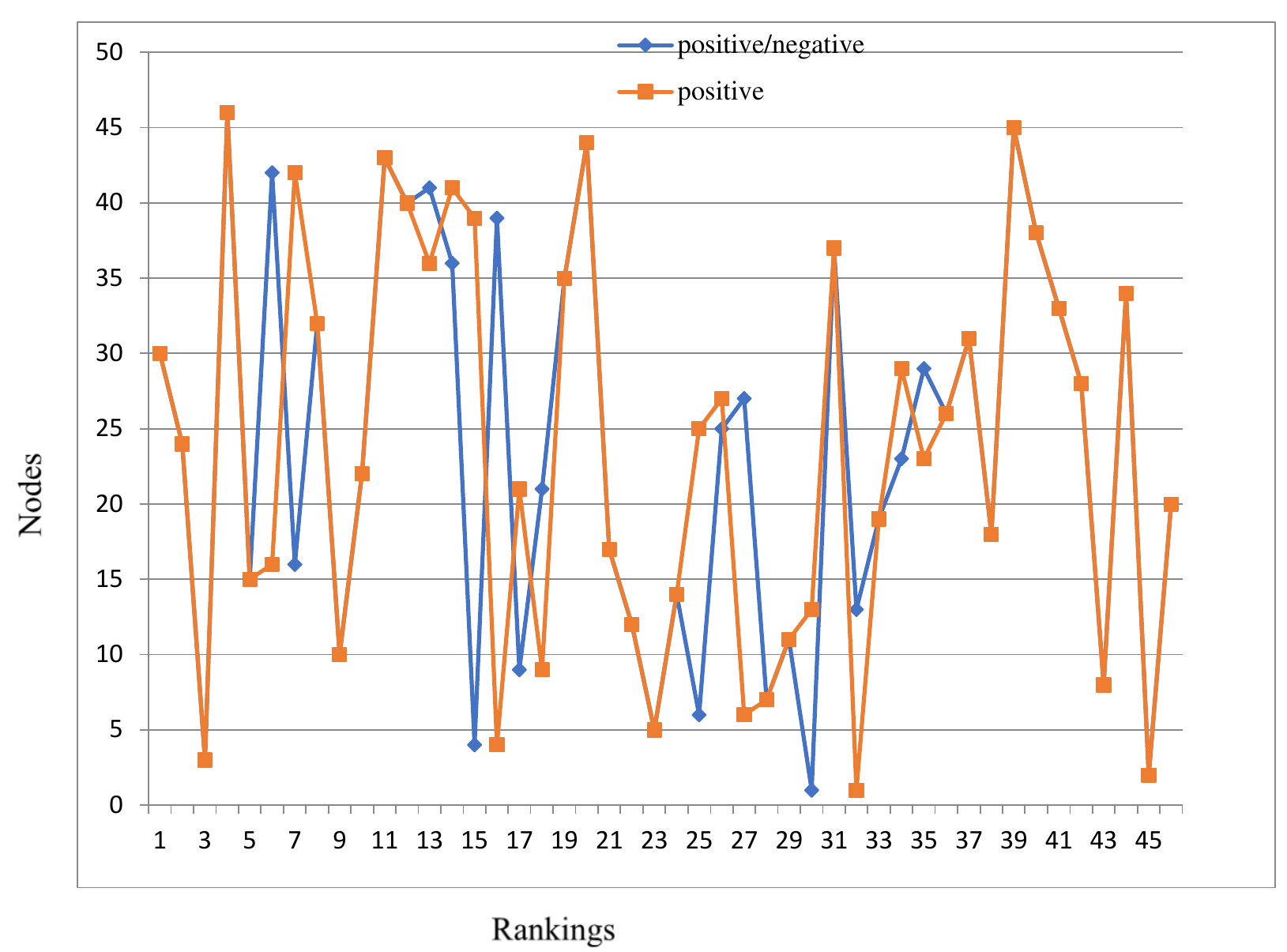}
  \caption{Interaction overlap graph DS1}
  \label{fig4}
\end{figure}

The second test was conducted to show the novice capability of the proposed EER technique to identify the difference between negative and positive interactions thereby solving the issue of negative referrals of graph-based techniques. For this DS1, DS2 was utilized. The data was manipulated by changing the weights of the expert node identified through the EER technique. The expert node identified through the BL1 technique was node 3. Both EER and BL1 algorithms were executed on the manipulated dataset. Since the EER technique can identify positive and negative interaction, so this time instead of node1 a new node is identified as an expert node i.e. node2. The BL1 algorithm however ranked the same node3 as an expert, this is because the BL1 algorithm is based upon the PageRank that cannot take into account the case of negative connections and referrals. It is pertinent to mention that none of the expert identification techniques have addressed this problem. This shows the ability of EER in identifying the experts by taking into account interactions categories, whereas the BL1 algorithm is unable to establish the categories of interactions. 
The authors utilized DS1 and DS2 datasets to carry out the experiments.
In one scenario nodes were ranked considering only positive interactions, whereas in the second scenario nodes were ranked by taking into account both positive and negative interactions, that were categorized by using a cutoff. The variance of top three nodes form both lists were calculated to find how close they reflect the original opinions.The results revealed that  values are closer to mean when both positive and negative interactions are utilized.
The table~\ref{t5} shows the MAE of the ranked list in above discussed two scenarios. Figure~\ref{fig4}shows that in case of DS1 very few instances overlap for both ranked lists.

 \begin{table}
\centering
\caption{MAE of EER, BL Algorithms}
\label{t5}
\scalebox{0.85}{
\begin{tabular}{@{}c c c c @{}}
\toprule
Avg Weight set1(+ve/-ve)&	Avg Weight set2(+ve)	&MAERank set1	&MAE Rank set2\\
\midrule
3.94	&3.60	&	0.14	& 0.20\\
3.93	&3.00	& 0.12  &	0.80\\
3.89	&3.72 &	0.07	& 0.09\\
\bottomrule
\end{tabular}}
\end{table}

\begin{figure}
\centering
  \includegraphics[width=1.0\linewidth]{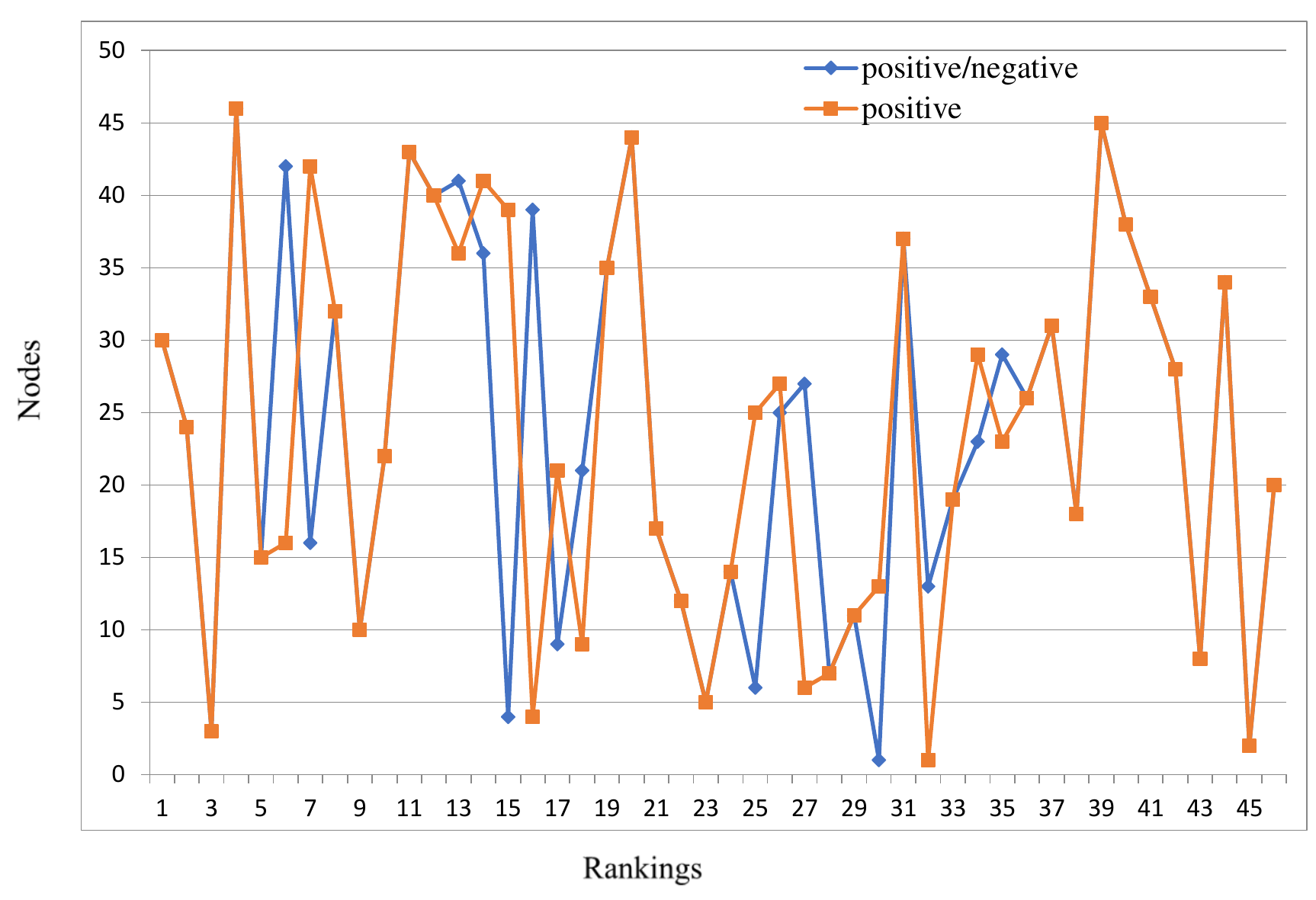}
  \caption{Interaction overlap graph DS2}
  \label{fig5}
\end{figure}

The same experiment when conducted on the data set DS2 showed almost 60\% overlap. To analyse further manual checking of the data set was carried out, revealing that nearly all of the nodes had positive interactions since the ratings were 3 or above.
Thus, the difference in the result was minimum. The graphical representation in Figure~\ref{fig5}depicts interaction overlap of two lists for DS2.

\subsection{Dynamic Behavior Test}

In order to verify the dynamism of the proposed technique EER, the authors took a specific scenario of a node with a history of 10 interactions with different other nodes. Using the proposed EER  with the time factor if $t=0$, it implies the usage of all the interactions. This produced an expected value of 0.4, as opposed to when t=1 i.e. only recent interaction is utilized as 0.3. 

\begin{figure}
\centering
  \includegraphics[width=0.95\linewidth]{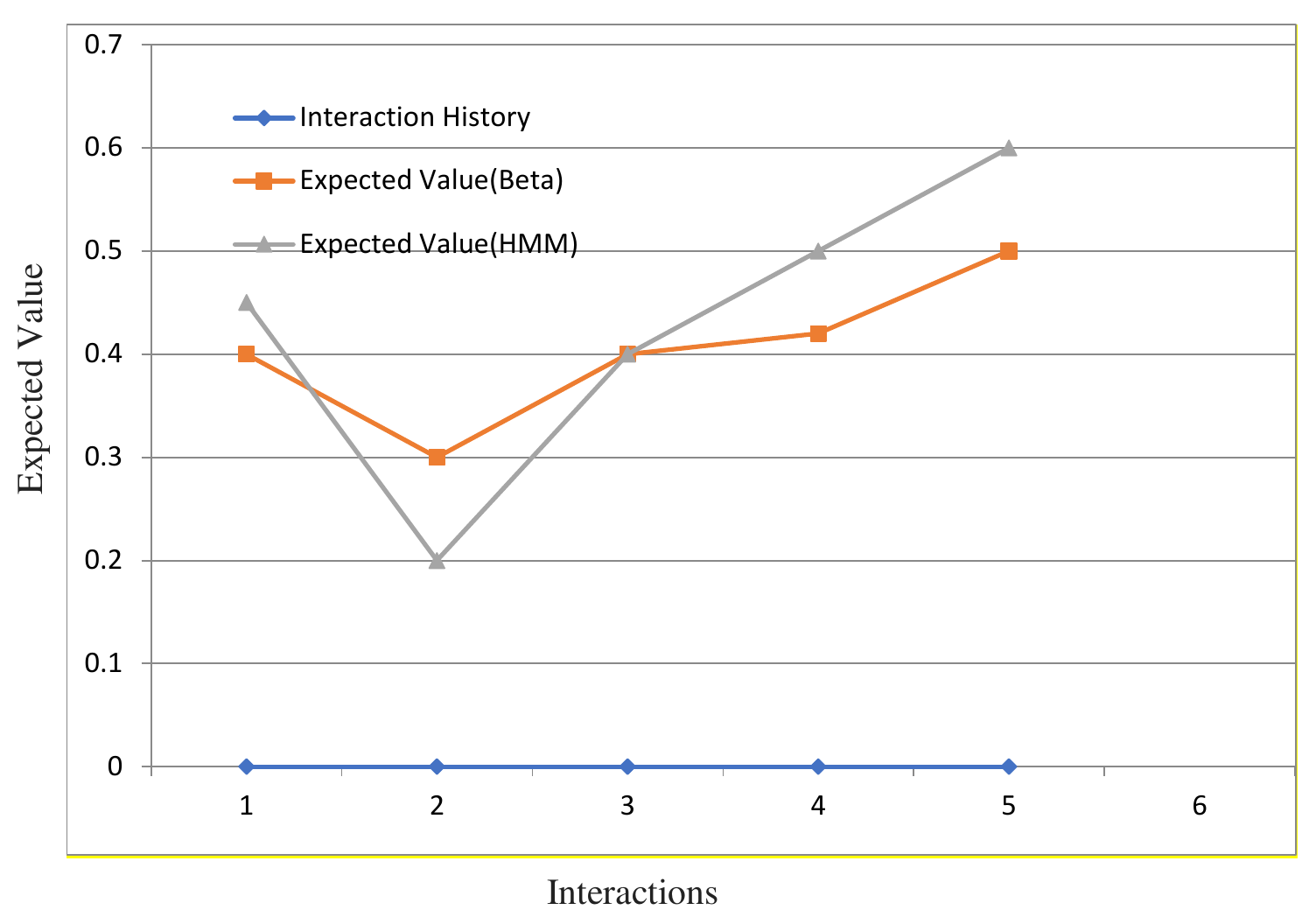}
  \caption{The hidden Markov reputation model and Beta reputation model with different observation probabilities.}
  \label{fig6}
\end{figure}

Given below in Table~\ref{t6} and Figure~\ref{fig6} is the representation of the expected value of a node with a variable history of interactions. Thus the problem of reputation staticity~\cite{4} is solved as addressed in the HMM-based technique. The authors are of the view that the HMM technique is a generalized version of beta distribution where each state individually assumed beta probability. From the literature, it is also evident that the HMM technique has limitations regarding state duration that follows geometric pattern unsuitable for real-life examples. Also, the number of hidden states needs to be declared apriori.In the authors view the HMM-based reputation model has added more complexity when the learning module is added.Furthermore,the working of the HMM model is restricted by parameter estimation that increases its complexity~\cite{31}. The model is hard to interpret for daily users, compared to the simple beta probability that is only parameterised by two variables. By closely observing the result of the proposed technique and HMM-based technique it is evident that to observe a change in behavior, the change in HMM-based technique is steeper comparatively, that shows its ability to respond to changing environment. The given scenario of this research is the capability of finding tacit experts. The tacit expertise usually do not suffer such rapid changes. Such changes are usually utilized to estimate the malicious behavior of the entity that previously  had a well-established behavior.

\begin{table}
\centering
\caption{Expected Value (observation probabilities) with varying history}
\label{t6}
\scalebox{0.85}{
\begin{tabular}{@{}l c c c @{}}
\toprule
Interaction History	& Expected Value(EER)	& Expected Value(HMM)\\
\midrule
All History&	0.40 & 0.45\\
Latest		 &	0.30 & 0.20\\
Latest 3	 &	0.40 & 0.40\\
Latest 5	 &	0.42 & 0.50\\
Latest 7	 &	0.50 & 0.60\\
\bottomrule
\end{tabular}}
\end{table}

\section{Conclusion}

Expert ranking techniques have recently gained lot of attention and research due to emergence of online knowledge communities, microblogs, internet based business markets. Most of the work in this domain targeted microblogs, however few took the same problem in the domain of large organizations and enterprises that are spread over continents. With the global spread of organizations, employees from diverse knowledge backgrounds join in, thus making expert identification a need for the managers, especially for the cases when identification of undocumented and tacit expertise of an employee is required. Thus the authors have proposed a reputation based technique to address the problem that has strong basis of mathematics and statistics. The proposed technique has addressed issue left unaddressed by previous techniques in terms of negative referral, collusion in case of graph based techniques. Comparison is conducted against domain specific techniques utilizing normal distribution reputation with its own limitations and HMM based reputation model. The EER technique is able to update according to the history of interactions so that only recent interactions could be utilized for calculations. The experimental results reveal better performance of EER in comparison to the three baselines in terms of MAE, Precision. 
 The EER technique is not restricted by the domain like previous techniques that were specifically designed for micro-blogs or email communication etc.It can be adopted for other domains as well. Furthermore with the emergence of global enterprises, and online job portals, such technique can also be helpful to fulfill the requirement of assignment of related task to employee with related expertise.

As a future enhancement, the application of the proposed scheme to other application domains will also be explored.

\bibliographystyle{cas-model2-names}
\bibliography{cas-refs}

\end{document}